\documentclass[12pt,twoside]{article}
\usepackage{amssymb}
\usepackage{amsmath}
\usepackage[dvips]{graphicx}
\begin{document}

\begin{center}
{\LARGE  Classical simulation of two spin-$S$ singlet state
correlations involving spin measurements.}\\
~\\

\vspace{0.4cm}
Ali Ahanj \footnote{Electronic address: ahanj@physics.unipune.ernet.in} ,  Pramod S. Joag\footnote{Electronic address: pramod@physics.unipune.ernet.in} \\

\vspace{0.2cm}
{\it Department of Physics, University of Pune,  Pune - 411007, India.}

\vspace{0.5cm}
Sibasish Ghosh \footnote{Electronic address: sibasish@imsc.res.in}\\

\vspace{0.2cm}
{\it The Institute of Mathematical Sciences, C. I. T.
Campus, Taramani, Chennai - 600 113, India.}
%\twocolumn
\end{center}

\begin{abstract} We give a classical protocol to exactly simulate
quantum correlations implied by a spin-$s$ singlet state for the
infinite sequence of spins satisfying $(2s + 1) = 2^{n}$, in the worst-case scenario, where $n$
is a positive integer. The class of measurements we consider here
are only those corresponding to spin observables. The required
amount of communication is found to be $log_{2}d$ where $d = 2s + 1$
is the dimension of the spin-$s$ Hilbert space.
\end{abstract}
%\maketitle

\vspace{0.3cm}
\begin{flushright}
\end{flushright}

%\end{abstract}
%\end{center}

\section{Introduction}
It is well known that quantum correlations implied by
an entangled quantum state of a bipartite quantum system cannot be
produced classically, i.e., using only the local and realistic
properties of the subsystems, without any communication between the
two subsystems [1]. By quantum correlations we mean the
statistical correlations between the outputs of measurements
independently carried out on each of the two entangled
parts. Naturally, the question arises as to the minimum amount of
classical communication (number of cbits) necessary to simulate the
quantum correlations of an entangled bipartite system. This amount of
communication quantifies the nonlocality of the entangled bipartite
quantum system. It also helps us gauge [2] the amount of
information hidden in the entangled quantum system itself in some
sense, the amount of information that must be space-like transmitted, in a local hidden variable model, in order for nature to account for the excess quantum correlations.

In this scenario, Alice and Bob try and output $\alpha$ and $\beta$
respectively, through a classical protocol, with the same probability
distribution as if they shared the bipartite entangled system and
each measured his or her part of the system according to a given
random Von Neumann measurement. As we have mentioned above, such a
protocol must involve communication between Alice and Bob, who
generally share finite or infinite number of random variables. The
amount of communication is quantified [3] either as the
average number of cbits $\overline {C}(P)$ over the directions along
which the spin components are measured (average or expected
communication) or the worst case communication, which is the maximum
amount of communication $C_{w}(P)$ exchanged between Alice and Bob
in any particular execution of the protocol. The third method is
asymptotic communication i.e., the limit
$lim_{n\rightarrow\infty}\overline{C}(P^{n})$ where $P^{n}$ is the
probability distribution obtained when $n$ runs of the protocol
carried out in parallel i.e., when the parties receive $n$ inputs
and produce $n$ outputs in one go. Note that, naively, Alice can just
tell Bob the direction of her measurement to get an exact classical
simulation, but this corresponds to  an infinite amount of
communication. the question whether a simulation can be done with
finite amount of communication was raised independently by
Maudlin [4], Brassard, Cleve and Tapp [5]
and Steiner [6]. Brassard, Cleve and
Tapp used the worst case communication cost while Steiner used the
average. Steiner's model is weaker as the amount of communication in
the worst case can be unbounded although such cases occur with zero
probability. Brassard, Cleve and Tapp gave a protocol to simulate
entanglement in a singlet state (i.e., the EPR pair) using eight
cbits of communication. Csirik [7] has improved it
where one requires six bits of communication. Toner and Bacon [8] 
gave a protocol to simulate two-qubit singlet state
entanglement using only one cbit of communication. Interestingly,
quantum correlations that cannot be classically simulated without
communication also occur in a scenario where incompatible
observables are successively measured on class of input (single
particle) spin-s states which can be simulated with a classical
protocol with communication between successive measurements [9].

Classical simulation of quantum correlations is
accomplished for spin-$1/2$ singlet state, requiring the optimal amount, namely, 1 cbit
of classical communication in the worst-case scenario, using arbitrary projective measurement on each site [8]. It is important to know
how does the amount of this classical communication change with the
change in the value of the spin $s$, in order to quantify the
advantage offered by quantum communication over the classical one.
Further, this communication cost quantifies, in terms of classical
resources, the variation of the nonlocal character of quantum
correlations with spin values. In this paper we give a classical
protocol to simulate the measurement correlation in a singlet state
of two spin-$s$ systems, considering only measurement of spin
observables (i.e., measurement of observables of the form
${\hat{a}}.{\vec{\Lambda}}$ where $\hat{a}$ is any unit vector in
${I\!\!\!R}^3$ and $\vec{\Lambda} = ({\Lambda}_x, {\Lambda}_y,
{\Lambda}_z)$ with each ${\Lambda}_i$ being a $(2s + 1) \times (2s +
1)$ traceless Hermitian matrix and the all three together form the
$SU(2)$ algebra) where $2s + 1 = 2^{n}$, $n$ being any positive
integer. As $s$ can only take integral or half-integral values,
the allowed values of $s$ form the infinite sequence
$s_{n = 1} = 1/2$, $s_{n = 2} = 3/2$, $s_{n = 3} = 7/2$, etc. For
$s_n ={2^{n-1} - 1}/{2}$, our protocol requires $n$ cbits of
communication and $2n$ number of independent and uniformly
distributed shared random variables.

We describe the general classical simulation of singlet state of two spin-$s$ systems in section II, followed by the specification of the measurement scenario, which we will be considering in the present paper. We  briefly describe the protocol of Toner and Bacon [8] in section III. We present our results for spin-$s$ singlet state in section IV. Finally, our conclusions are summarized in section V. 

\section{Correlation of two spin-$S$ singlet state}
The singlet state $|{\psi}^-_s\rangle_{AB}$ of two spin-$s$
particles $A$ and $B$ is the eigenstate corresponding to the
eigenvalue $0$ of the total spin observable of these two spin
systems, namely the state
\begin{equation}
\label{singlet} |{\psi}^-_s\rangle_{AB} = \frac{1}{\sqrt{2s + 1}} \sum_{m =
-s}^{s} (- 1)^{s - m} |m\rangle_A \otimes |- m\rangle_B,
\end{equation}
where $|- s\rangle$, $|- s + 1\rangle$, $\ldots$, $|s - 1\rangle$,
$|s\rangle$ are eigenstates of the spin observable of each of the
individual spin-$s$ system. Thus $|{\psi}^-_s\rangle_{AB}$ is a
maximally entangled state of the bipartite system $A + B$, described
by the Hilbert space ${\mathbb{C}}^{2s + 1} \otimes {\mathbb{C}}^{2s
+ 1}$. In the case of classical simulation of the quantum
correlation $\langle{\psi}^-_{1/2}|\hat{a}.\vec{\sigma} \otimes
\hat{b}.\vec{\sigma}|{\psi}^-_{1/2}\rangle$ of the two-qubit singlet
state $|{\psi}^-_{1/2}\rangle$, Alice considers measurement of
traceless observable $\hat{a}.\vec{\sigma}$ and Bob considers that
of the traceless observable $\hat{b}.\vec{\sigma}$ . These are spin
observables. Analogous to the Pauli matrices, one can consider $(2s
+ 1)^2 - 1$ number of trace-less but trace-orthogonal Hermitian $(2s + 1) \times
(2s + 1)$ matrices ${\Lambda}_1$, ${\Lambda}_2$, $\ldots$,
${\Lambda}_{(2s + 1)^2 - 1}$ ({\it i.e.}, ${\rm Tr} {\Lambda}_i = 0$ for all $i$ but ${\rm Tr} ({\Lambda}_i {\Lambda}_j) = 0$ if $i \ne j$; see, for example,[10] )
such that a general projective measurement on the individual
spin-$s$ system corresponds to the measurement of an observable of
the form $\hat{c}.\vec{\Lambda}$, where $\hat{c}$ is a unit vector
in ${I\!\!\!R}^{(2s + 1)^2 - 1}$ and $\vec{\Lambda}$ is the $((2s + 1)^2 - 1)$-tuple $({\Lambda}_1, {\Lambda}_2, \ldots, {\Lambda}_{(2s + 1)^2 - 1})$ of the above-mentioned $\Lambda$ matrices. In general, the quantum
correlation $\langle{\psi}^-_s|\hat{c}.\vec{\Lambda} \otimes
\hat{d}.\vec{\Lambda}|{\psi}^-_s\rangle$ will be a bilinear function
in the components on $\hat{c}$ and $\hat{d}$. In the special
case when $({\Lambda}_i \otimes {\Lambda}_j) |{\psi}^-_s\rangle =
{\alpha} {\delta}_{ij} |{\psi}^-_s\rangle$ for all $i, j = 1, 2,
\ldots, (2s + 1)^2 - 1$, the quantum correlation will be of the
form ${\alpha} \hat{c}.\hat{d}$ . 

%Tracelessness of the observable
%$\hat{c}.\vec{\Lambda}$ is important as its expectation value with
%the reduced density matrix $\frac{1}{2s + 1} I$ of the individual
%system (when the joint state is the singlet state
%$|{\psi}^-_s\rangle$) should vanish as in the case of spin-1/2
%singlet.

Classical simulation of this general quantum correlation seems to be
quite hard one possible reason being the absence of Bloch sphere
structure for higher spin systems. Rather we will consider only
measurement of spin observables, namely the observables of the form
$\hat{a}.{\bf J}$ on each individual spin-$s$ system, where
$\hat{a}$ is an arbitrary unit vector in ${I\!\!\!R}^3$ and ${\bf J}
= (J_x, J_y, J_z)$. For the $(2s + 1) \times (2s + 1)$ matrix
representations of the spin observables $J_x$, $J_y$, and $J_z$,
please see  page 191 - 192 of ref.[11].

$J$ matrices satisfy the $SU(2)$ algebra, namely $[J_x, J_y] =
iJ_z$, $[J_y, J_z] = iJ_x$, $[J_z, J_x] = iJ_y$. The eigenvalues of
$\hat{a}.{\bf J}$ are $- s$, $- s + 1$, $\ldots$, $s - 1$, $s$ for
all $\hat{a} \in {I\!\!\!R}^3$. The quantum correlations
$\langle{\psi}^-_s|\hat{a}.{\bf J} \otimes \hat{b}.{\bf
J}|{\psi}^-_s\rangle$ (which we will denote here as $\left\langle
\alpha \beta\right\rangle$, where $\alpha$ runs through all the
eigenvalues of $\hat{a}.{\bf J}$ and $\beta$ runs through all the
eigenvalues of $\hat{b}.{\bf J}$) is given by
\begin{equation}
\label{correlation} \langle{\psi}^-_s|\hat{a}.{\bf J} \otimes
\hat{b}.{\bf J}|{\psi}^-_s\rangle = \left\langle \alpha
\beta\right\rangle=-\frac{1}{3} s(s+1)\hat{a}.\hat{b} \;,
\end{equation}
where $\hat{a}$ and $\hat{b}$ are the unit vectors specifying the
directions along which the spin components are measured by Alice and
Bob respectively [12]. Note that, by virtue of being a
singlet state ,$\left\langle \alpha \right\rangle = 0 =
\left\langle\beta\right\rangle$ irrespective of directions $\hat{a}
$ and $\hat{b}$. From now onward, we will consider only those
spin-$s$ systems for which $2s + 1 = 2^n$, $n$ being any positive
integer. Thus we see that the allowed spin systems will form the
sub-class $\{2^{n - 1} - 1/2 : n = 1, 2, \ldots\}$ of half-integral
spins.

\section{Classical simulation of two spin-1/2 singlet state}
As the working principles of our protocol are of similar in nature
with those of Toner and Bacon [8], before describing our
protocol, we would like to briefly describe the protocol of Toner
and Bacon to simulate the measurement correlations on
$|{\psi}^-_{1/2}\rangle$. In this scenario, Alice and Bob's job is
to simulate the quantum correlation
$\langle{\psi}^-_{1/2}|\hat{a}.{\frac{1}{2}}{\mathbf{\sigma}}
\otimes \hat{b}.{\frac{1}{2}}{\mathbf{\sigma}}|{\psi}^-_{1/2}\rangle
= - \frac{1}{4}\hat{a}.\hat{b}$, together with the conditions that
$\langle \alpha \rangle = 0 = \langle \beta \rangle$. To start with,
Alice and Bob share two independent random variables $\hat{\lambda}$
and $\hat{\mu}$, each of which has uniform distribution on the
surface of the Bloch sphere $S_2$ in ${I\!\!\!R}^3$. Given the
measurement direction $\hat{a}$, Alice calculates $-
\frac{1}{2}sgn(\hat{a}.\hat{\lambda})$, which she takes as her measurement
output $\alpha$. Note that $sgn(x) = 1$ for all $x \ge 0$ and
$sgn(x) = - 1$ for all $x < 0$. As $\hat{\lambda}$ is uniformly
distributed on $S_2$, for each given $\hat{a}$, $-
\frac{1}{2}sgn(\hat{a}.\hat{\lambda})$ will take its values $\frac{1}{2}$ and $- \frac{1}{2}$ with
equal probabilities, i.e., ${\rm Prob} (\alpha = 1/2) =~ {\rm Prob}
(\alpha = - 1/2) = 1/2$ (and hence, $\langle \alpha \rangle = 0$).
Alice then sends the one bit information $c \equiv
sgn(\hat{a}.\hat{\lambda}) sgn(\hat{a}.\hat{\mu})$ to Bob. Note that
instead of sending $sgn(\hat{a}.\hat{\lambda})$, by sending $c$,
Alice does not allow Bob to extract any information about her output
$\alpha$. This is so because ${\rm Prob} (\alpha = 1/2 | c = 1) =~
{\rm Prob} (\alpha = - 1/2 | c = 1)$ and ${\rm Prob} (\alpha = 1/2 | c =
- 1) =~ {\rm Prob} (\alpha = - 1/2 | c = - 1)$. After receiving $c$,
and using his measurement direction $\hat{b}$, Bob now calculates
his output $\beta \equiv \frac{1}{2}sgn[\hat{b}.(\hat{\lambda} + c
\hat{\mu})]$. Now
\begin{equation}
\label{betavalue}
\langle \beta \rangle = \frac{1}{2(4\pi)^2} \int_{\hat{\lambda} \in
S_2}~ \int_{\hat{\mu} \in S_2}~ sgn[\hat{b}.(\hat{\lambda} +
sgn(\hat{a}.\hat{\lambda})sgn(\hat{a}.\hat{\mu}) \hat{\mu})]
d\hat{\lambda} d\hat{\mu}\;.
\end{equation}
Given any $\hat{\mu} \in S_2$, for each
choice of $\hat{\lambda} \in S_2$, the two values of the integrand
corresponding to $\hat{\lambda}$ and $- \hat{\lambda}$ are negative
of each other. As the distribution of $\hat{\lambda}$ on $S_2$ is
taken to be uniform, the above-mentioned observation
immediately shows that $\langle \beta \rangle = 0$. As $\beta \in
\{1/2, - 1/2\}$, therefore ${\rm Prob} (\beta = 1/2) =~ {\rm Prob} (\beta
= - 1/2) = 1/2$. In order to compute $\langle \alpha \beta \rangle$,
one should observe that Bob's output can also be written as $\beta =
\frac{1}{2}\sum_{d = \pm 1} [(1 + cd)/2] sgn[\hat{b}.(\hat{\lambda} + d \;
\hat{\mu})]$. The following two among the four integrals (which
appears in $\langle \alpha \beta \rangle$)
$$- \frac{1}{8(4\pi)^2} \int_{\hat{\lambda} \in S_2}~
\int_{\hat{\mu} \in S_2}~ sgn(\hat{a}.\hat{\lambda})
sgn[\hat{b}.(\hat{\lambda} \pm \hat{\mu})] d\hat{\lambda}
d\hat{\mu}$$ cancels each other by incorporating the inversion
$\hat{\mu} \rightarrow - \hat{\mu}$. And the rest two integrals
$$\pm \frac{1}{8(4\pi)^2}  \int_{\hat{\lambda} \in S_2}~
\int_{\hat{\mu} \in S_2}~ sgn(\hat{a}.\hat{\lambda})
sgn[\hat{b}.(\hat{\lambda} \pm \hat{\mu})] d\hat{\lambda}
d\hat{\mu}$$ are same and they are equal to the integral
$$\frac{1}{8(4\pi)^2}  \int_{\hat{\lambda} \in S_2}~
\int_{\hat{\mu} \in S_2}~ sgn(\hat{a}.\hat{\lambda})
sgn[\hat{b}.(\hat{\mu} - \hat{\lambda})] d\hat{\lambda}
d\hat{\mu}\;.$$ And hence we have
$$\langle{\psi}^-_{1/2}|\hat{a}.{\mathbf{\sigma}}
\otimes \hat{b}.{\mathbf{\sigma}}|{\psi}^-_{1/2}\rangle \equiv
\langle \alpha \beta \rangle =$$ 
\begin{equation}
\label{alphabeta}
-\frac{2}{8(4\pi)^2}
\int_{\hat{\lambda} \in S_2}~ \int_{\hat{\mu} \in S_2}~
sgn(\hat{a}.\hat{\lambda}) sgn[\hat{b}.(\hat{\mu} - \hat{\lambda})]
d\hat{\lambda} d\hat{\mu} = - \frac{2}{8(4\pi)} \int_{\hat{\lambda} \in
S_2}~ sgn(\hat{a}.\hat{\lambda}) \hat{b}.\hat{\lambda} d\hat{\lambda} = -
\frac{1}{4}\hat{a}.\hat{b}\;.
\end{equation}

\section{Classical simulation of two spin-$S$ singlet state using spin measurements} 
Let us now come to our protocol. In the simulation of the
measurement of the observable $\hat{a}.{\bf J}$ (where $\hat{a} \in
{I\!\!\!R}^3$ is the supplied direction of measurement), Alice will
have to reproduce the $2^n$ number of outcomes $\alpha = 2^{n - 1} -
1/2, 2^{n - 1} - 3/2, \ldots, - 2^{n - 1} + 1/2$ with equal
probability. If we consider the series $-\frac{1}{2} \sum_{k =
1}^{n} f(k) 2^{n - k}$, where, for each $k$, $f(k)$ can be either
$1$ or $- 1$, it turns out that the series can only take the
above-mentioned $2^n$ different values of $\alpha$. The
probability distribution of these different values of the series
will depend on that of the $n$-tuple $\{f(1), f(2), \ldots, f(n)\}$.
In order to make this probability distribution an uniform one (which
is essential here for the simulation purpose), we choose here $f(k)
= sgn(\hat{a}.{\hat{\lambda}}_k)$ for each $k$, where $\hat{a}$
is the measurement direction for Alice while
${\hat{\lambda}}_1$, ${\hat{\lambda}}_2$, $\ldots$,
${\hat{\lambda}}_n$ are independent and uniformly distributed
random variables on $S_2$. We have seen in the above-mentioned Toner
and Bacon protocol that if Alice and Bob share the two independent
and uniformly distributed random variables ${\hat{\lambda}}_k
\in S_2$ and ${\hat{\mu}}_k \in S_2$, then the random variable
$r_k \equiv sgn[\hat{b}.({\hat{\lambda}}_k +
sgn(\hat{a}.{\hat{\lambda}}_k) sgn(\hat{a}.{\hat{\mu}}_k)
{\hat{\mu}}_k)]$ is uniformly distributed over $\{1, - 1\}$.
Hence, as above, the quantity $\frac{1}{2} \sum_{k = 1}^{n} 2^{n -
k} sgn[\hat{b}.({\hat{\lambda}}_k + r_k {\hat{\mu}}_k)]$
will have $2^n$ different values $\beta = 2^{n - 1} - 1/2, 2^{n - 1}
- 3/2, \ldots, - 2^{n - 1} + 1/2$ all with equal probabilities. But
the interesting point to note is that in the calculation of the
average (over the independent but uniformly distributed random
variables ${\hat{\lambda}}_1$, ${\hat{\lambda}}_2$,
$\ldots$, ${\hat{\lambda}}_n$, ${\hat{\mu}}_1$,
${\hat{\mu}}_2$, $\ldots$, ${\hat{\mu}}_n$) of the product
$\alpha \beta$, there will be no contribution from cross terms like
$sgn(\hat{a}.{\hat{\lambda}}_k)
sgn[\hat{b}.(\hat{{\lambda}_l} + r_l \hat{{\mu}_l})]$ if $k
\ne l$.

The protocol proceeds as follows: Alice outputs
$\alpha=-\frac{1}{2}\sum^{n}_{k=1}2^{n-k}sgn(\hat{a}.{\hat{\lambda}}_k)$.
Alice sends $n$ cbits $c_{1}, c_{2}, \dots, c_{n}$ to Bob where
$c_{k} =
sgn(\hat{a}.{\hat{\lambda}}_k)sgn(\hat{a}.{\hat{\mu}}_k)$
for $k = 1, 2, \ldots, n$, where ${\hat{\lambda}}_1$,
${\hat{\lambda}}_2$, $\ldots$, ${\hat{\lambda}}_n$,
${\hat{\mu}}_1$, ${\hat{\mu}}_2$, $\ldots$,
${\hat{\mu}}_n$ are independent shared random variables
between Alice and Bob, each being uniformly distributed on $S_2$. Thus we see that, in terms of shared randomness, $\lambda = ({\hat{\lambda}}_1, {\hat{\lambda}}_2, \ldots, {\hat{\lambda}}_n, {\hat{\mu}}_1, {\hat{\mu}}_2, \ldots, {\hat{\mu}}_n)$ is the shared random variable between Alice and Bob. After receiving these $n$ cbits from Alice, Bob outputs $\beta =
\frac{1}{2}\sum^{n}_{k = 1} 2^{n -
k}sgn[\hat{b}.({\hat{\lambda}}_k + c_{k} {\hat{\mu}}_k)]$.
It follows immediately from the discussion in the last paragraph
that
$$\langle \alpha \beta \rangle = -\frac{1}{4}\frac{1}{(4\pi)^{2n}}\sum^{n}_{k = 1}2^{2n - 2k}
\int d{\hat{\lambda}}_1 \ldots d{\hat{\lambda}}_{k -
1}d{\hat{\lambda}}_{k + 1} \ldots
d{\hat{\lambda}}_nd{\hat{\mu}}_1 \ldots d{\hat{\mu}}_{k -
1}d{\hat{\mu}}_{k + 1} \ldots d{\hat{\mu}}_n \times $$
\begin{equation}
\label{finalcor1} \int_{({\hat{\lambda}}_k, {\hat{\mu}}_k)
\in S_2 \times S_2} sgn(\hat{a}.{\hat{\lambda}}_k)
sgn[\hat{b}.({\hat{\lambda}}_{k} + d_{k}{\hat{\mu}}_{k})]
d{\hat{\lambda}}_k d{\hat{\mu}}_k.
\end{equation}
It follows from the discussion in section III regarding Toner and
Bacon's work that $\langle \alpha \beta \rangle = -
\frac{1}{4}\sum^{n}_{k = 1} 2^{2n - 2k} \hat{a}.\hat{b}$ . Summing
the geometric series and using $(2s + 1) = 2^{n}$ we finally get
\begin{equation}
\label{finalform} \langle \alpha \beta \rangle = - \frac{1}{3} s (s
+ 1) \hat{a}.\hat{b}\;.
\end{equation}

This protocol exactly simulates quantum mechanical probability
distribution for particular types of projective measurements, namely
the spin measurement, on the spin $s$ singlet state with $2s + 1 =
2^{n}$ for positive integer $n$. The above protocol applies to
infinite, although sparse, subset of the set of all spins (i.e., all
integral and half integral values). The most important finding is
that the amount of communication goes as $log_{2}(2s + 1)$ or as
$log_{2} s$ for $s \gg 1$. Our protocol works equally for any two
spin-$s$ maximally entangled state as that can be locally unitarily
connected to the singlet state.

\section{Conclusion}
Our result provides the amount of classical communication in the
worst case scenario if we consider only measurement of spin
observables on both sides of a two spin-$s$ singlet state with the
restriction that the dimension $2s + 1$ of each subsystem must be a
positive integral power of $2$, and just $n = log_2 (2s + 1)$ bits of
communication from Alice to Bob is sufficient. We are unable to show whether our protocol is optimal (in the sense of using minimum number of classical communication).   
On the other hand, if we consider most general projective measurements on
both the sides of a maximally entangled state of two qudits, with $d
= 2^n$, it is known that (see [5]) Alice would require
at least of the order of $2^n$ bits of communication to be sent to
Bob, in the worst case scenario when $n$ is large enough. But for
general $d$, $log_2 d$ can be shown to be a lower bound on the
average amount of classical communication that one would require to
simulate the maximally entangled correlation of two qudits
considering most general type of projective measurements
[13]. It is also known that $log_2 d$ bits of classical
communication on average is sufficient to simulate the measurement
correlation of a maximally entangled state of two qudits, when both
Alice and Bob consider only measurement of traceless binary
observables [14]. It thus seems that even if simulation
of maximally entangled correlation in the most general case of
projective measurement is a hard problem, and one would require to
send classical communication at least of the order of the dimension
(for large dimensional case), there is still some room to search
for efficient simulation protocols in lower dimensions.

\vspace{0.3cm}
{\textbf{Acknowledgment:}} We thank Guruprsad Kar and R. Simon for encouragement. A.A. and P.S.J. thank Andr\'e M\'ethot for an effective correspondence. S.G. thanks the Physics Department, Pune University for its hospitality during which part of this work was done. Finally, we would like to thank the anonymous referee for valuable comments and suggestions to revise the earlier version of the present manuscript.

\end{document}